\documentclass[aps,prl,twocolumn,superscriptaddress]{revtex4-1}
\usepackage{graphicx}
\usepackage{epstopdf}
\usepackage{dcolumn}
\usepackage{bm}
\usepackage{subfigure}
\usepackage{amsmath}
\usepackage{epsfig}
\usepackage{verbatim}
\usepackage{float}
\usepackage{color}
\usepackage{url}

\begin{document}
\title{Impacts of export restrictions on the global personal protective equipment trade network during COVID-19}
\author{Yang Ye}
\affiliation{School of Data Science, City University of Hong Kong, Hong Kong SAR, China}

\author{Qingpeng Zhang}
\email{qingpeng.zhang@cityu.edu.hk}
\affiliation{School of Data Science, City University of Hong Kong, Hong Kong SAR, China}

\author{Zhidong Cao}
\affiliation{The State Key Laboratory of Management and Control for Complex Systems, Institute of Automation, Chinese Academy of Sciences, Beijing, China}
\affiliation{School of Artificial Intelligence, University of Chinese Academy of Sciences, Beijing, China}
\affiliation{Shenzhen Artificial Intelligence and Data Science Institute (Longhua), Shenzhen, China}

\author{Frank Youhua Chen}
\author{Houmin Yan}
\affiliation{Department of Management Sciences, College of Business, City University of Hong Kong, Hong Kong SAR, China}

\author{H. Eugene Stanley}
\affiliation{Department of Physics, Boston University, Boston, MA 02215, USA}

\author{Daniel Dajun Zeng}
\affiliation{The State Key Laboratory of Management and Control for Complex Systems, Institute of Automation, Chinese Academy of Sciences, Beijing, China}
\affiliation{School of Artificial Intelligence, University of Chinese Academy of Sciences, Beijing, China}
\affiliation{Shenzhen Artificial Intelligence and Data Science Institute (Longhua), Shenzhen, China}

\begin{abstract}
The COVID-19 pandemic has caused a dramatic surge in demand for personal protective equipment (PPE) worldwide. Many countries have imposed export restrictions on PPE to ensure the sufficient domestic supply. The surging demand and export restrictions cause shortage contagions on the global PPE trade network. Here, we develop an integrated network model, which integrates a metapopulation model and a threshold model, to investigate the shortage contagion patterns. The metapopulation model captures disease contagion across countries. The threshold model captures the shortage contagion on the global PPE trade network. Results show that, the shortage contagion patterns are mainly decided by top exporters. Export restrictions exacerbate the shortages of PPE and cause the shortage contagion to transmit even faster than the disease contagion. Besides, export restrictions lead to ineffective and inefficient allocation of PPE around the world, which has no benefits for the world to fight against the pandemic.
\end{abstract}

\maketitle
The COVID-19 pandemic is spreading rapidly around the world. As of Jan 24, 2021, it has infected more than 96 million people and claimed more than 2 million lives worldwide \cite{WHO1, li2020early}. Many countries have adopted a series of public health measures to contain the epidemic, such as the closure of commercial activities, bans on travel, and export restrictions \cite{anderson2020will,he2020short,laborde2020covid}. Personal protective equipment (PPE), such as face and eye protection devices, protective garments, and gloves, is the most heavily affected category of commodities in export restrictions. Over 73 governments have imposed export restrictions on PPE exports \cite{wtoexportrestriction,bown2020covid}. Since the COVID-19 pandemic has caused a growing demand for PPE worldwide \cite{burki2020global,rowan2020challenges,park2020global}, countries impose export restrictions to prepare for the potential domestic demand.

Recently, several empirical studies have discussed the pros and cons of export restrictions on medical supplies, foods, drugs, etc., in the time of COVID-19. They concluded that export restrictions might cause uncertainty in supply and other negative security consequences, though these restrictions seem logical and justifiable \cite{laborde2020covid,bown2020covid,pauwelyn2020export}. Demand surges and export restrictions cause shortage contagion on the trade network. There is rich economic literature using quantitative models to investigate the contagion patterns and their impacts on international trade \cite{guan2020,wenz2016enhanced,burkholz2019international,gephart2016vulnerability,distefano2018shock}. In physics, a wide range of research proposed different models to analyze the dynamics of contagion propagation on interdependent networks \cite{buldyrev2010catastrophic}, for example, the diffusion model \cite{contreras2014propagation} and the threshold model \cite{watts2002simple, centola2007cascade}. However, how the shortage contagion transmits on the global PPE trade network during large-scale epidemic like the ongoing COVID-19 pandemic is under-researched. Most, if not all, existing studies examined the disease contagion and shortage contagion separately, and did not take into consideration the dynamic interplay between them. It is critical to characterize such interplay because the surging demand is caused by the epidemic arrival.

\begin{figure}[htbp]
\includegraphics[width=1.0\columnwidth]{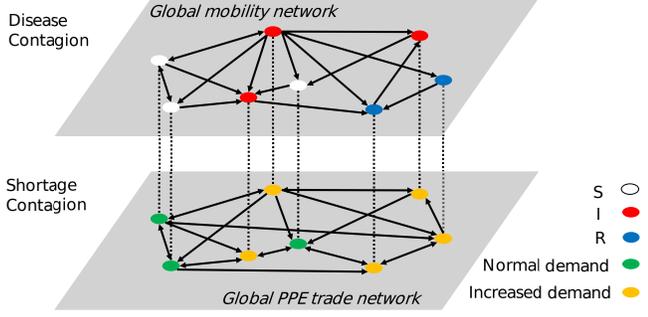}
\caption{\label{multi}Overview of the integrated network model. The top (disease contagion) and bottom (shortage contagion) layers are the global mobility network where the epidemic spreads, and the global PPE trade network where the shortage contagion transmits, respectively. Nodes represent countries. Edges on the top layer represent the aggregated number of seats on scheduled commercial flights between countries per day. Edges on the bottom layer represent the daily trade value between countries (in US dollars). Countries' domestic demand for PPE will increase since they are infected.}
\end{figure}

In this paper, we develop a novel integrated network model to examine the impacts of export restrictions on the global PPE trade network during the COVID-19 pandemic. We illustrate the structure of the model in Fig.~{\ref{multi}}. The proposed model integrates a susceptible-infected-recovered (SIR) based metapopulation model, which captures the dynamics of disease contagion  on the global mobility network (top layer of Fig.~{\ref{multi}}), and a threshold model, which captures the dynamics of shortage contagion on the global PPE trade network (bottom layer of Fig.~{\ref{multi}}). We investigate the shortage propagation patterns of eight sections of PPE commodities for five scenarios on export restrictions. We provide quantitative evidence that export restrictions cause shortage contagion to transmit even faster than that of the disease contagion. Besides, export restrictions delay the occurrence of shortages for self-sufficient countries, but accelerate the occurrence of shortages for not-self-sufficient countries. In addition, export restrictions lead to ineffective and inefficient allocation of PPE worldwide.

To capture the interplay between the disease contagion and the demand for PPE, we adapt a threshold model \cite{burkholz2019international, watts2002simple} by (a) representing the increase of domestic demand for PPE as a result of the epidemic arrival, and (b) adding an inventory module as the buffering mechanism. We construct the global PPE trade network using data from the United Nations Comtrade Database (UNCD) \cite{uncomtrade2018}. Here nodes represent countries and edges represent the annual trade value between countries (in US dollars). We select commodities with the World Customs Organization's Harmonized System codes for COVID-19 medical supplies \cite{WCO2020}. All commodities are classified into eight sections. 

Following Brockmann and Helbing \cite{brockmann2013hidden}, the SIR based metapopulation model is constructed based on the global mobility network, which is defined by the daily air traffic data \cite{oag2013}. Here, nodes represent countries and edges represent the aggregated number of seats on scheduled commercial flights between countries. The population data used for constructing the global mobility network is obtained from the United Nations World Population Prospects national estimates \cite{popu2013}. After excluding the countries that do not appear in these datasets, the proposed model has 195 countries. 

The model works on a daily time step. At time $t$, each country $i$ distributes its imports $imp_i^{(s)}(t)$, production $pro_i^{(s)}(t)$, and inventory $inv_i^{(s)}(t)$ of commodities in section $s$ to meet the domestic demand $dem_{i,dom}^{(s)}(t)$ and foreign demand $dem_{i,for}^{(s)}(t)$. We assume that domestic demand has higher priority than foreign demand. The maximum amount of commodities in section $s$ that country $i$ can distribute at time $t$ is
\begin{equation}
    D_{i,max}^{(s)}(t)=imp_i^{(s)}(t)+pro_i^{(s)}(t)+inv_i^{(s)}(t).
\end{equation}
If the epidemic arrives in country $i$ at time $t$, the domestic demand for commodities in all sections will increase since then. Denote $N_i$ as the population size of country $i$, and $S_i(t)$, $I_i(t)$, and $R_i(t)=N_i-S_i(t)-I_i(t)$ as the number of susceptible, infected, and recovered individuals at time $t$, respectively. Then, following \cite{brockmann2013hidden}, the dynamics of disease contagion is given by
\begin{equation}
\begin{aligned}
\partial_tI_i(t) &= \frac{\alpha S_i(t)I_i(t)\sigma(\frac{I_i(t)}{\epsilon N_i})}{N_i}-\beta I_i(t)\\
&+\gamma\sum_{j\neq i}P_{ji}[I_j(t)-I_i(t)],\\
\partial_tS_i(t) &= -\frac{\alpha S_i(t)I_i(t)\sigma(\frac{I_i(t)}{\epsilon N_i})}{N_i}\\
&+\gamma\sum_{j\neq i}P_{ji}[S_i(t)-S_j(t)].
\label{sir}
\end{aligned}
\end{equation}
Here, $\alpha$, $\beta$, $\gamma$, and $\epsilon$ are the infection rate, recovery rate, average mobility rate, and local invasion parameter, respectively. $\sigma(x)=x^{\eta}/(x^{\eta}+1)$ is the sigmoid function with parameter $\eta$. Denote $F_{ji}$ as the number of passengers traveling from country $i$ to country $j$ per day, and $F_i=\sum_{j}F_{ji}$. Then $P_{ji}=F_{ji}/F_i$ is the fraction of individuals traveling from country $i$ to country $j$, and $\sum_{j}P_{ji}=1$. We obtain $F_{ji}$ by averaging the daily traffic data on the global mobility network. 

Assuming that, before the pandemic, the domestic demand for commodities in section $s$ is $\mu_i^{(s)}$ per capita per day for country $i$, thus, the total domestic demand for commodities in section $s$ for country $i$ before the pandemic is $Dem_i^{(s)}=\mu_i^{(s)}N_i$ per day. We adopt the common assumption that the domestic demand for PPE increases with the number of confirmed cases, and then reaches a plateau. To capture this relationship, we modify the relationship function in \cite{Worby2020} and represent $dem_{i,dom}^{(s)}(t)$ as 
\begin{equation}
\begin{split}
    dem_{i,dom}^{(s)}(t)=&Dem_i^{(s)}[1+\theta_{d,i}(t)]\\
    =&Dem_i^{(s)}+\theta_{d,i}(t)\mu_i^{(s)}N_i.
\end{split}
\label{dem increase}
\end{equation}
Here, $\theta_{d,i}(t)$ is the demand increase factor for country $i$ at time $t$, which is represented as follows.
\begin{equation}
    \theta_{d,i}(t) = k_1\Bigg\{\frac{2}{1 + e^{-k_2\big[1-\frac{S_i(t)}{N_i}\big]}}-1\Bigg\},
\end{equation}
where $k_1>0$ quantifies the upper limit of $\theta_{d,i}(t)$ and $k_2>0$ quantifies the level of ``panic buying" effect (i.e. consumers buy unusually large amounts of PPE commodities in anticipation of, or after, the epidemic arrival). Country $i$'s domestic demand (for commodities in section $s$) fulfilled by $i$ itself can be expressed as 
\begin{equation}
    dem_{i,dom,a}^{(s)}(t) = \text{min}\{D_{i,max}^{(s)}(t), dem_{i,dom}^{(s)}(t)\}.
\end{equation}
Without export restrictions, the maximum foreign demand for commodities in section $s$ to be fulfilled by country $i$ can be expressed as 
\begin{equation}
\begin{aligned}
        dem_{i,for, max}^{(s)}(t) ={}& \text{min}\{D_{i,max}^{(s)}(t)-dem_{i,dom, a}^{(s)}(t), \\
        &dem_{i,for}^{(s)}(t)\}.
\end{aligned}
\end{equation}
Denote the proportion of commodities in section $s$ being exported from country $i$ to country $j$ as
\begin{equation}
    x_{i,j}^{(s)} = \frac{W_{i,j}^{(s)}}{Exp_{i}^{(s)}},
\end{equation}
and we assume that $x_{i,j}^{(s)}$ is constant. Here, $W_{i,j}^{(s)}$ is the amount of commodities in section $s$ that country $i$ exports to country $j$ before the pandemic, and $Exp_{i}^{(s)}$, country $i$'s total exports of commodities in section $s$ is 
\begin{equation}
    Exp_{i}^{(s)}=\sum_jW_{i,j}^{(s)}.
\label{exports}
\end{equation}
$W_{i,j}^{s}$ is obtained from the UNCD. Then, we can derive the actual amount of commodities that country $i$ exports to country $j$ at time $t$ as
\begin{equation}
    w_{i,j}^{(s)}(t) = x_{i,j}^{(s)}dem_{i,for, max}^{(s)}(t)r_{i,j},
\end{equation}
where $r_{i,j}\in\{0,1\}$, and $r_{i,j}=0$ if country $i$ restricts exports to country $j$; otherwise $r_{i,j}=1$. Thus,
\begin{equation}
    imp_j^{(s)}(t+1)=\sum_iw_{i,j}^{(s)}(t),
\end{equation}
and the inventory of commodities in section $s$ that country $i$ holds at the beginning of the next period (i.e., the end of the current period) is 
\begin{equation}
    inv_{i}^{(s)}(t+1) = D_{i,max}^{(s)}(t)-dem_{i,dom, a}^{(s)}(t)-\sum_jw_{i,j}^{(s)}(t).
\end{equation}
A lower inventory level than the initial level will result in an increase in production, thus, the production at the next period is decided as follows.
\begin{equation}
pro_{i}^{(s)}(t+1)=
\begin{cases}
Pro_i^{(s)}& inv_i^{(s)}(t)\geq inv_i^{(s)}(0),\\
Pro_i^{(s)}[1+\theta_{p,i}(t)] & \text{otherwise},
\end{cases}
\label{pro increase}
\end{equation}
where $\theta_{p,i}(t)$ is the production increase factor for country $i$ at time $t$. We assume that $\theta_{p,i}(t)$ is non-negative for the following reasons. During the pandemic, PPE production may decline due to the lockdown of cities, infection of workers, etc. But in the meanwhile, governments have provided supports for PPE production and manufacturers worldwide have retooled to produce more PPE to combat the pandemic. Therefore, we assume the production after the pandemic is no less than that before the pandemic.

Assuming that countries cannot anticipate economic shocks, they issue orders to other countries at the end of each time period based on $pro_{i}^{(s)}(t+1)$, $dem_{i,dom}^{(s)}(t)$, and $dem_{i,for}^{(s)}(t)$. The total amount of commodities in section $s$ that country $i$ orders from other countries is 
\begin{equation}
\begin{split}
    imp_{i,o}^{(s)}(t) =& \text{max}\{dem_{i,dom}^{(s)}(t)+dem_{i,for}^{(s)}(t)\\
    &-pro_{i}^{(s)}(t+1), 0\}.
\end{split}
\end{equation}
Denote the proportion of commodities in section $s$ being imported from country $j$ to country $i$ as
\begin{equation}
    y_{j,i}^{(s)} = \frac{W_{j,i}^{(s)}}{Imp_{i}^{(s)}},
\end{equation}
and we assume that $y_{j,i}^{(s)}$ is constant. Here, the total amount of commodities in section $s$ that country $i$ imports from other countries is
\begin{equation}
    Imp_{i}^{(s)}=\sum_jW_{j,i}^{(s)}.
\label{imports}
\end{equation}
 Therefore, the amount of commodities that country $i$ orders from country $j$ is $y_{j,i}^{(s)}imp_{i,o}^{(s)}(t)$, and
\begin{equation}
    dem_{j,for}^{(s)}(t+1) = \sum_iy_{j,i}^{(s)}imp_{i,o}^{(s)}(t).
\end{equation}
We assume that, before the pandemic, 
\begin{equation}
    Imp_i^{(s)}+Pro_i^{(s)}=Exp_i^{(s)}+Dem_i^{(s)}.
\label{trans}
\end{equation} 
We initialize the model by setting $pro_i^{(s)}(0)=Pro_i^{(s)}$, $imp_i^{(s)}(0)=Imp_i^{(s)}$, and $dem_{i,for}^{(s)}(0)=Exp_i^{(s)}$.
We assume that 
\begin{equation}
    inv_i^{(s)}(0)=\phi_i^{(s)}Imp_i^{(s)}.
\end{equation}
This assumption means that country $i$ can still meet the domestic demand and foreign demand without imports for $\phi_i^{(s)}$ days before the pandemic. 

In the simulations, we consider the simplest pandemic scenario, where no travel bans or other public health measures are considered. For simplicity, we assume $k_1=2$, $k_2=100$, and $\mu_i^{s}=10$ for all countries and all commodities. Following epidemiology literature, the mean infectious period is set as 4.6 days \cite{moghadas2020projecting} leading to the recovery rate $\beta=0.217$. The basic reproduction number $R_0$ is set as 2.6 \cite{riou2020pattern}, leading to the infection rate $\alpha=R_0\beta=0.5642$. From the OAG data, the average mobility rate $\gamma$ is estimated to be $0.00014$ per day. We adopt the choices for $\epsilon$ and $\eta$ in \cite{brockmann2013hidden}, $\epsilon=10^{-8}$ and $\eta=8$. We set that China is initially infected with 100 infected cases at $t=0$, which corresponds to December 31, 2019, the date when the World Health Organization was informed of unknown pneumonia cases detected in Wuhan, China \cite{whotime}. We run the simulation for one year.

\begin{table}
\caption{\label{scenarios}Description of five export restriction scenarios.}
\begin{ruledtabular}
\begin{tabular}{cc}
Scenario&Description\\
\colrule
$\text{S}_{none}$ & No country restricts exports\\
$\text{S}_{1}$ & The largest exporter restricts exports\\
$\text{S}_{5\%}$ & The top 5\% of exporters restrict exports\\
$\text{S}_{lower}$ & The lower half (50\%) of exporters restrict exports\\
$\text{S}_{all}$ & All countries restrict exports
\end{tabular}
\end{ruledtabular}
\end{table}

Now, we model five different export restriction scenarios among countries, and present their impacts on the trade network for each commodity section. The description of each scenario is given in Table~\ref{scenarios}. For the rest of this paper, we only present the numerical results for commodities in section 1 (COVID-19 test kits and apparatus used in diagnostic testing) and section 2 (protective garments and the like), because they represent two typical situations: (a) the initially infected country (China) is not the largest exporter (Germany) in section 1, and (b) the initially infected country (China) is also the largest exporter in section 2. Results for other sections are consistent with the results for these two sections, and thus are presented in the Supplemental Material. 

\begin{figure}[htbp]
\includegraphics[width=1.0\columnwidth]{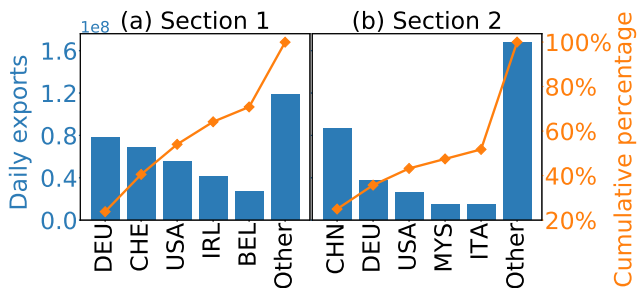}
\caption{\label{Pareto}The Pareto distribution in global exports. The daily exports (left) and the cumulative percentage of daily global exports (right) in (a) section 1 and (b) section 2 for the top five exporters and other countries.  DEU = Germany, CHE = Switzerland, USA = the United States of America, IRL = Ireland, BEL = Belgium, CHN = China, MYS = Malaysia, ITA = Italy.}
\end{figure}

First, we give an overview of the trade network for section 1 and section 2. In Fig.~\ref{Pareto}, we present the daily exports and the cumulative percentage of daily global exports in section 1 and section 2 for the top five exporters and other countries. We observe a Pareto distribution in global exports in Fig.~\ref{Pareto}, where the top five exporters share about 70\% and 52\% of global exports in section 1 and section 2, respectively.

\begin{figure}[htbp]
\includegraphics[width=1.0\columnwidth]{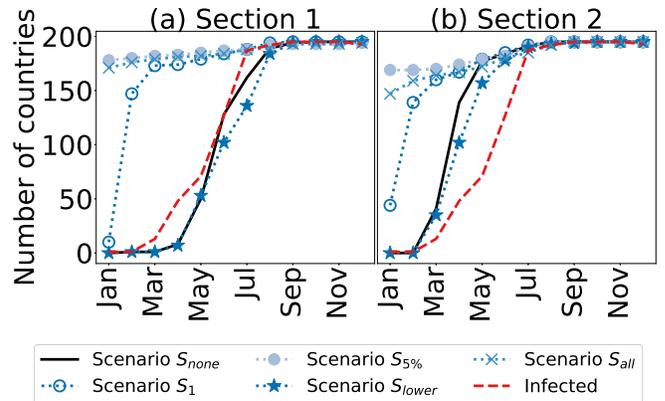}
\caption{\label{infected_num}The number of infected countries and the number of countries facing shortages of commodities in (a) section 1 and (b) section 2 at the end of each month. The description of scenarios is given in Table~\ref{scenarios}. Parameters are set as follows: $\theta_{p,i}(t)=0$, $\phi_i^{(s)}=10$.}
\end{figure}

In Fig.~\ref{infected_num}, we plot the number of infected countries and the number of countries facing shortages at the end of each month in different scenarios. Country $i$ will face shortages of commodities in section $s$ at time $t$ if its domestic demand cannot be met, i.e., $D_{i,max}^{(s)}(t)<dem_{i,dom}^{(s)}(t)$. As illustrated in Fig.~\ref{infected_num}, if the production stays unchanged for all countries (i.e., $\theta_{p,i}(t)=0$), nearly all countries will face shortages at the end of 2020 in all scenarios. Generally, export restrictions exacerbate global supply shortages. We can observe that, compared with scenario $S_{all}$ where all countries restrict exports, the number of countries facing shortages decreases greatly in the early periods (from January to June) in scenario $S_{none}$ where no country restricts exports. Besides, the number of countries facing shortages in scenario $S_{5\%}$ (only the top 5\% of exporters restrict exports) is nearly the same as scenario $S_{all}$, which can be explained by the Pareto distribution in global exports. Counterintuitively, the number of countries facing shortages in scenario $S_{lower}$ (the lower half of the exporters restrict exports) is slightly fewer than that in scenario $S_{none}$. The reason is as follows. The total percentages of world exports that the lower half of the exporters share are only 0.005\% and 0.053\% for section 1 and section 2, respectively. They can hardly help other countries manage supply shortages. Besides, exports lower the inventory level of these countries, thus, in scenario $S_{lower}$, restricting exports can delay the shortages when disease arrives at these countries. These findings show that, on the one hand, agreements are urgently needed to ensure open trade between countries during the pandemic; on the other hand, such agreements should also allow countries that contribute little to global exports to impose some export restrictions in order to ensure a sufficient inventory for the upcoming pandemic.

For section 1, we find that scenario $S_{1}$ leads to many more countries facing shortages as compared to scenario $S_{none}$, because Germany (the largest exporter) stops supplying commodities to others. This gap shrinks for section 2, because the largest exporter (China) is the initially infected country as well. So even without export restrictions (such as scenario $S_{none}$), China has to meet the domestic demand by lowering the exports significantly. Moreover, we find that for both sections, there are more countries facing shortages than infected countries in scenarios $S_{1}$, $S_{5\%}$, and $S_{all}$. This finding indicates that pandemic-resulted export restrictions can make shortage contagion transmit even faster than disease contagion.

\begin{figure}[htbp]
\includegraphics[width=1.0\columnwidth]{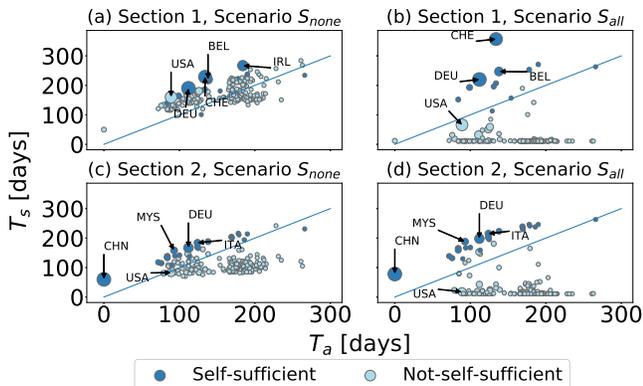}
\caption{\label{tats}Comparison of the epidemic arrival time $T_a$ and the first shortage time $T_s$ for each country in (a) section 1, scenario $S_{none}$, (b) section 1, scenario $S_{all}$, (c) section 2, scenario $S_{none}$, and (d) section 2, scenario $S_{all}$. Nodes represent countries. The size of a node represents the export value of commodities in the corresponding section. The color of a node indicates if it is a self-sufficient country for the section. The blue line corresponds to $T_a=T_s$. Parameters are set as follows: $\theta_{p,i}(t)=0$, $\phi_i^{(s)}=10$. For clarity, only the top five exporters are presented with the three-letter country code. DEU = Germany, CHE = Switzerland, USA = the United States of America, IRL = Ireland, BEL = Belgium, CHN = China, MYS = Malaysia, ITA = Italy.}
\end{figure}
Next, we compare the epidemic arrival time $T_a$ and the first shortage time $T_s$ for each country in different scenarios. We present the results for scenarios $S_{none}$ and $S_{all}$ in Fig.~\ref{tats}. Note that countries not facing shortages within the simulation periods are not represented in Fig.~\ref{tats}. The epidemic arrival time $T_a$ is defined as the date of the first infected case after the initial outbreak. The first shortage time $T_s$ is defined as the date when a country first faces PPE shortages. Nodes represent countries. The size of a node represents the export value of commodities in the corresponding section. The color of a node indicates if it is a self-sufficient country for the section. Country $i$ is
self-sufficient for commodities in section $s$ when its production is no less than its domestic demand before the pandemic, i.e., $Pro_i^{(s)}>=Dem_i^{(s)}$. The blue line corresponds to $T_a=T_s$. If a country faces shortages before infected, it locates below the blue line. Otherwise, it locates above the blue line. 

\begin{figure}[htbp]
\includegraphics[width=1.0\columnwidth]{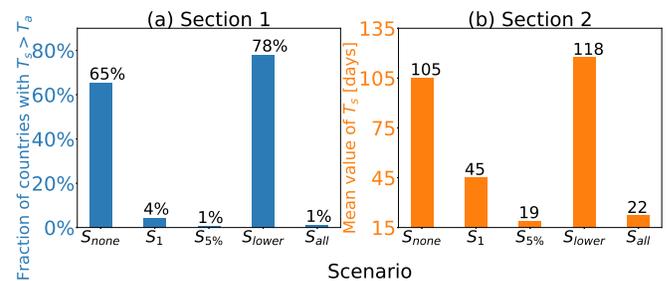}
\caption{\label{tsta_1} (a) Fraction of not-self-sufficient countries with $T_s>T_a$ in section 1 and (b) the mean value of $T_s$ in section 2 for not-self-sufficient countries in all scenarios. Parameters are set as follows: $\theta_{p,i}(t)=0$, $\phi_i^{(s)}=10$.}
\end{figure}

As illustrated in Fig.~{\ref{tats} (a)} and Fig.~{\ref{tats} (b)}, we can observe in section 1 that, compared with scenario $S_{all}$, more not-self-sufficient countries locate above the blue line in scenario $S_{none}$. The fraction of not-self-sufficient countries above the blue line increases to 65\% in scenario $S_{none}$ from 1\% in scenario $S_{all}$. In section 2, we also observe that export restrictions lead to a much earlier occurrence of shortages for not-self-sufficient countries in Fig.~{\ref{tats} (c)} and Fig.~{\ref{tats} (d)}. The mean value of $T_s$ decreases to 22 days in scenario $S_{all}$ from 105 days in scenario $S_{none}$. In Fig.~{\ref{tsta_1}}, we also present the fraction of countries with $T_s>T_a$ in section 1 and the mean value of $T_s$ in section 2 for not-self-sufficient countries in all scenarios. The differences in scenarios are consistent with that in Fig.~{\ref{infected_num}}. To sum up, when all countries restrict exports, almost all not-self-sufficient countries locate below the blue line, which can be observed in both section 1 and section 2, indicating that they encounter PPE shortages even before the epidemic arrival. Besides, both self-sufficient countries and not-self-sufficient countries locate farther away from each other in scenario $S_{all}$ than in scenario $S_{none}$. Similar results to scenario $S_{all}$ are found in scenario $S_{1}$ and scenario $S_{5\%}$. These results indicate that export restrictions delay the occurrence of shortages for self-sufficient countries, but accelerate the occurrence of shortages for not-self-sufficient countries.

These results also present the double-edged nature of the PPE trade relationship between countries. On the one hand, such relationships allow countries (especially not-self-sufficient countries) to address the pandemic with the help of trading partners. On the other hand, shortage contagion can also transmit through these relationships. As illustrated in Fig.~\ref{tats}(c), 80 countries suffer from shortages while only 26 countries are infected at $t=100$. The surge in domestic demand in infected countries leads to a reduction in their exports. The shortage contagion then spills over to non-infected countries because the domestic demand cannot be met. However, countries can mitigate such spillover effects by increasing production while facing shortages. We present the averaged fraction of countries facing shortages for each month with different production increase factor $\theta_{p,i}(t)$ in scenarios $S_{none}$ and $S_{all}$ in the Supplemental Material. We observe in both sections that, the number of countries facing shortages decreases greatly as $\theta_{p,i}(t)$ grows, especially at the early stage (from January to June). Besides, even as $\theta_{p,i}(t)$ grows, there are still more countries facing shortages in scenario $S_{all}$ than that in scenario $S_{none}$ for the same $\theta_{p,i}(t)$. These results indicate that cooperation between countries (no export restrictions) always plays an essential role in preventing global shortages of PPE regardless of the production level. But at the same time, a higher production level leads to less dependence on imports, which greatly helps countries cope with PPE shortages. Therefore, except for promoting global cooperation, governments and international organizations should take actions to reduce supply chain barriers and work together to increase global PPE production.

\begin{figure}[htbp]
\includegraphics[width=1.0\columnwidth]{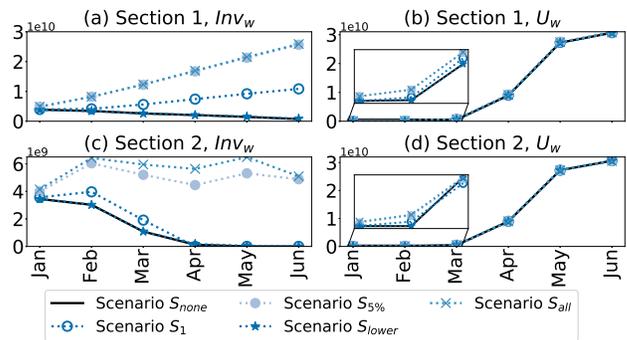}
\caption{\label{inv_dem}Average world total inventory $Inv_{w}^{(s)}(t)$ (a, c) and world total unmet domestic demand $U_{w}^{(s)}(t)$ (b, d) of section 1 (a, b) and section 2 (c, d) from January to June in all scenarios. Parameters are set as follows: $\theta_{p,i}(t)=0$, $\phi_i^{(s)}=10$}
\end{figure}

Finally, we compare the world total inventory $Inv_{w}^{(s)}(t)$ and the world total unmet domestic demand $U_{w}^{(s)}(t)$ for both sections at the end of time $t$. We define $Inv_{w}^{(s)}(t)$ = $\sum_{i}inv_i^{(s)}(t+1)$ and $U_{w}^{(s)}(t)$ = $\sum_{i}dem_{i,dom}^{(s)}(t)-dem_{i,dom,a}^{(s)}(t)$. We present the average values of $Inv_{w}^{(s)}(t)$ and $U_{w}^{(s)}(t)$ from January to June in Fig.~{\ref{inv_dem}}. If $Inv_{w}^{(s)}(t)>0$, the world total inventory level $Inv_{w}^{(s)}(t)$ and the world total unmet domestic demand $U_{w}^{(s)}(t)$ in scenarios $S_{none}$, $S_{1}$, and $S_{lower}$ are all lower than that in scenarios $S_{5\%}$ and $S_{all}$. Compared with scenario $S_{all}$, $U_{w}^{(s)}(t)$ in scenario $S_{none}$ is reduced by 100\%, 93\%, and 24\% for January, February, and March in section 1, respectively. In section 2, $U_{w}^{(s)}(t)$ is reduced by 100\%, 100\%, and 0.64\% for January, February, and March, respectively. These results show that, with export restrictions, a large amount of PPE is hoarded instead of being distributed to where it is most needed, particularly at the early stage. We can also find that, although there are more countries facing shortages in scenario $S_{5\%}$ than that in scenario $S_{all}$ (Fig.~{\ref{infected_num}}), $Inv_{w}^{(s)}(t)$ and $U_{w}^{(s)}(t)$ in scenario $S_{5\%}$ are almost the same as or even slightly lower than that in scenario $S_{all}$. From this perspective, we can conclude that the more top exporters restrict exports, the less effective the global PPE supply chain is. These findings further indicate that export restrictions are not an appropriate solution to address the pandemic. A fully functional PPE supply chain system could leave countries more time to adapt their production and identify alternative supply sources. Countries should lift the export restrictions to help allocate PPE more effectively and efficiently for the collective benefit of humankind.

In summary, we investigated how the shortage contagion, induced by demand surges and export restrictions, transmits on the global PPE trade network during the COVID-19 pandemic. We simulated the impacts of five export restriction scenarios based on an integrated network model, which integrates the real-world PPE trade data and global mobility data. We find evidence that the shortage contagion pattern is mainly determined by the export restriction policies of the top exporters. Export restrictions can cause shortage contagion to transmit even faster than the disease contagion, with only the top 5\% of exporters imposing export restrictions. To some extent, export restrictions can provide benefits for self-sufficient countries, at the sacrifice of immediate economic shocks at not-self-sufficient countries. The results also validate that export restrictions are not an effective and efficient solution to confront the pandemic. PPE is not properly allocated to countries with shortages. To better respond to the next wave of COVID-19 and other emerging infectious diseases, countries should keep PPE trade open and reduce reliance on only a small number of PPE exporters. 

%

\end{document}